\documentclass[twocolumn]{raa}
\usepackage{graphicx,times}
\usepackage{natbib}
\usepackage{amssymb,amsmath}
\bibpunct{(}{)}{;}{a}{}{,}

\usepackage[a4paper=true,driverfallback=dvipdfm,pagebackref=true]{hyperref}

\hypersetup{colorlinks = true, linkcolor = green, anchorcolor = red, citecolor = blue, filecolor = red, pagecolor = red, urlcolor = red}

\mathchardef\mhyphen="2D
\defcitealias{Henriques2015}{H15}

\begin{document}
 \title{The Formation and Evolution of Massive Galaxies}

 \volnopage{ {\bf 20XX} Vol.\ {\bf X} No. {\bf XX}, 000--000}
   \setcounter{page}{1}

\author{Yingjie Jing\inst{1,2}, Yu Rong\inst{3}\thanks{FONDECYT postdoctoral fellow}, Jie Wang\inst{1,2}, Qi Guo\inst{1,2}, Liang Gao\inst{1,2,4}
      }


   \institute{ Key Laboratory for Computational Astrophysics, National Astronomical Observatories, Chinese Academy of Sciences, Beijing 100012, China; {\it jyj@nao.cas.cn}\\
         \and
             School of Astronomy and Space Science, University of Chinese Academy of Sciences, Beijing 10049, China\\
	     \and
       Instituto de Astrof\'isica, Pontificia Universidad Cat\'olica de Chile, Av. Vicu\~na Mackenna 4860, Macul, Santiago, Chile; {\it rongyuastrophysics@gmail.com}  \\
    \and 
   Institute of Computational Cosmology, Department of Physics, University of Durham, Science Laboratories, South Road, Durham DH1 3LE, UK\\
 \vs \no
   {\small Received 20XX Month Day; accepted 20XX Month Day}
}

\abstract{The discovery of massive galaxies at high redshifts, especially the passive ones, poses a big challenge for the current standard galaxy formation models. Here we use the semi-analytic galaxy formation model developed by Henriques et al. to explore the formation and evolution of massive galaxies (MGs, stellar-mass $M_{*}> 10^{11}$ M$_{\sun}$). Different from previous works, we focus on the ones just formed (e.g. just reach $\simeq 10^{11}$ M$_{\sun}$). We find that most of the MGs are formed around $z=0.6$, with the earliest formation at $z>4$. Interestingly, although most of the MGs in the local Universe are passive, we find that only $13\%$ of the MGs are quenched at the formation time. Most of the quenched MGs at formation already hosts a very massive supermassive black hole (SMBH) which could power the very effective AGN feedback. For the star-forming MGs, the ones with more massive SMBH prefer to quench in shorter timescales; in particular, those with $M_{\textrm{SMBH}} > 10^{7.5}$ M$_{\sun}$ have a quenching timescale of $\sim 0.5$ Gyr and the characteristic $M_{\textrm{SMBH}}$ depends on the chosen stellar mass threshold in the definition of MGs as a result of their co-evolution. We also find that the ``in-situ'' star formation dominates the stellar mass growth of MGs until they are formed. Over the whole redshift range, we find the quiescent MGs prefer to stay in more massive dark matter halos, and have more massive SMBH and less cold gas masses. Our results provide a new angle on the whole life of the growth of MGs in the Universe.
}

   \authorrunning{Jing et al.}            
   \titlerunning{The Formation and Evolution of Massive Galaxies}  
   \maketitle
%
\section{Introduction}
The studies based on semi-analytic galaxy formation models \citep[e.g.][]{Kauffmann1993,Baugh1996,DeLucia2006,DeLucia2007,Neistein2006,Guo2008,Parry2009,Lee2013} and hydrodynamical simulations \citep[e.g.][]{Oser2010,Lackner2012,Qu2016,Rodriguez-Gomez2016} show that the low-redshift massive galaxies may form most of stellar masses in their progenitor galaxies at high redshifts through in-situ star formation (50 per cent at $z\sim5$; \citet{DeLucia2007}), and then assemble through galaxy mergers later \citep[e.g.][]{Rong2018}. This formation scenario can explain the observational results that the most massive galaxies appear to have a large 
number of older stellar populations \citep[e.g.][]{Bower1992,Cowie1996,Heavens2004,Cimatti2004,Glazebrook2004,Gallazzi2005,Thomas2019}, and higher $\alpha$-element enhancement/shorter star-formation timescales \citep[e.g.][]{Thomas2010,Johansson2012,Zheng2019}, which is referred to as `downsizing' effect. However, the observational surveys \citep[e.g.][]{Mobasher2005, Wiklind2008} have also found a significant population of massive galaxies at high redshifts ($z>2$), supporting a drastic stellar mass growth in these galaxies at early epochs.

Most of the nearby massive galaxies are quiescent; besides, a noticeable fraction ($\sim 30$ per cent) of the massive galaxies have already been quenched at $z\simeq2$
\citep[e.g.][]{Daddi2005,Kriek2008,Fontana2009,Castro-Rodrguez2012,Nayyeri2014,Wang2016,Merlin2018,Deshmukh2018}. Particularly, recent studies report several extremely-high-redshift ($3\lesssim z\lesssim 4$) quiescent galaxies \citep{Glazebrook2017,Simpson2017,Schreiber2018,Valentino2020}, which provide a challenge to the current galaxy formation models. We note that it is essential to investigate the formation of these quiescent massive galaxies, at high-redshift particularly, as the current formation models are difficult to reconcile the required high star-formation rates (SFRs; e.g., $>100\ M_{\odot}$/yr) in the short star-formation timescales ($<1$~Gyr represented by the high [$\alpha$/Fe]; \citet{Kriek2016}) and thus drastic stellar mass growth, as well as efficient AGN/stellar feedback ceasing the violent star formation at early Universe.

The studies of \citet{Rong2017b} and \citet{Qin2017}, based on semi-analytic models, suggest that the AGN feedback indeed can cause the formation of the extremely-high-redshift (e.g., $z \sim 3.7$) massive quiescent galaxies, if their progenitors host more massive central black holes. \citet{Terrazas2016} also found that the quiescent galaxies host more massive black holes, comparing with the star-forming counterparts, with the data of 91 nearby massive galaxies ($z<0.034$), which also suggests a correlation between the central black hole masses and quiescence of galaxies.

In this work, we will present more details on the formation and evolution, particularly the quenching process, of the massive galaxies during the entire cosmic epoch by using the semi-analytic model developed by \citet{Henriques2015} based on the Millennium simulation \citep{Springel2005}. Semi-analytic galaxy formation models have been proven successful in reproducing the properties for both of the massive and dwarf galaxies in the local Universe and at high redshifts \citep[e.g.,][]{Guo2011, Guo2013, Yates2012, Buitrago2017, Rong2017a, Rong2017b}.
 
The organization of the paper is as follows. In Section 2, we briefly introduce the simulation and galaxy formation model used in this study. Then the formation and evolution of MGs and the properties at different redshift will be studied. Finally, the conclusions and discussions will be followed. Throughout this paper, we use 'log' to represent 'log$_{10}$'.

\section{Semi-analytic galaxy catalogue}

The galaxy catalogue we used is based on a semi-analytic galaxy formation model described in \citet[][hereafter \citetalias{Henriques2015}]{Henriques2015} implemented on the Millennium simulation \citep[MS;][]{Springel2005}. The MS adopts a flat $\Lambda$CDM cosmology model with cosmological parameters based on a combined analysis of the 2dFGRS \citep{Colless2001} and the first-year \textit{WMAP} data \citep{Spergel2003}. 
The full particle data were stored at 64 snapshots approximately logarithmically spaced from $z=20$ until $z=2$ and then at approximately 300 Myr intervals until $z=0$. At each snapshot, the particles were linked with a separation of less than 0.2 of the mean inter-particle separation \citep{Davis1985} to generate a friend-of-friend (FOF) group catalogue. Subsequently, the SUBFIND algorithm \citep{Springel2001} was applied to each FOF group to identify all its self-bounded subhaloes. 

\citetalias{Henriques2015} used the technique detailed in \citet{Angulo2010} and \citet{Angulo2015}  to scale the evolution of dark matter structure predicted by the MS to \textit{Planck} cosmology. The cosmological parameters adopted from Planck Collaboration XVI (\citeyear{PlanckCollaboration2014}) are $\Omega_{\rm{M}} = 0.315$, $\Omega_{\Lambda} = 0.685$, $\Omega_{\rm{b}} = 0.0487$ ($f_{\rm{b}} = 0.155$), $n = 0.96$, $\sigma_{\rm{8}} = 0.829$ and $H_{0} = 67.3$ km s$^{-1}$ Mpc$^{-1}$. After scaled, the simulation has a resolution of $2160^{3}$ particles in a periodic box of side length $480.279 h^{-1}$ Mpc with a particle mass of $9.6 \times 10^{8} h^{-1}$ M$_{\odot}$.  The box volume is large enough to investigate the statistical properties of massive galaxies.
 
\citetalias{Henriques2015} updated the previous  Munich models \citep{Guo2011,Guo2013} and provided a good fit to the observed stellar mass functions and passive fractions over the full redshift range, $0\leq z \leq 3$. This makes it possible to trace the evolution of the massive galaxies to high redshift.
 
 \begin{figure}
	\includegraphics[width=\columnwidth]{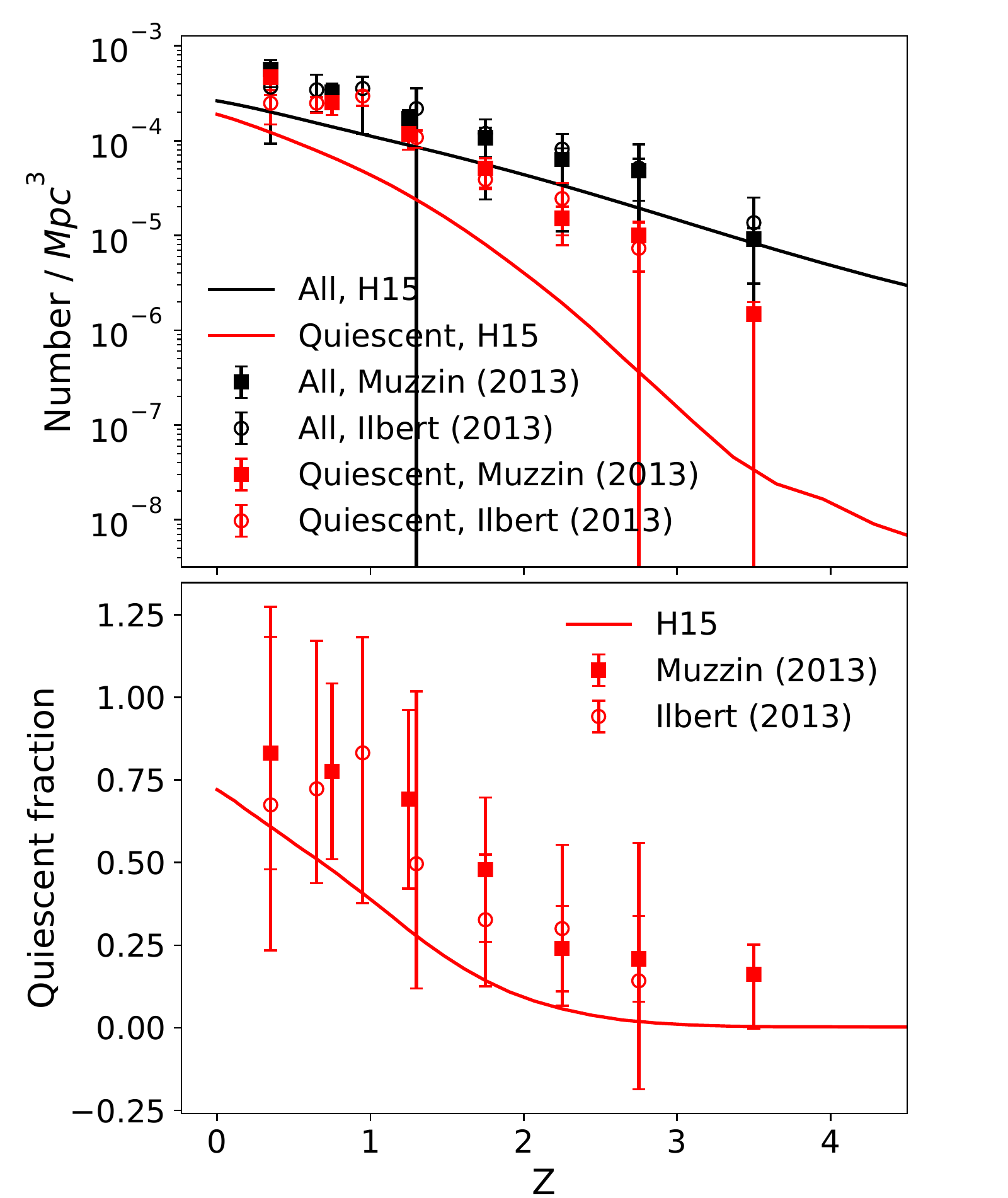}
	\caption{Number density (top panel) and quiescent fraction (bottom panel) of MGs (massive galaxies with stellar mass $M_* \geq 10^{11}$M$_{\sun}$) as a function of redshift. The solid lines show the result from \citetalias{Henriques2015} model which are calculated after the stellar masses been convolved with a Gaussian in log$M_{*}$, with width 0.08 $\times (1 + z)$, 
    The square points represent the observation data from  \citet{Muzzin2013}; The open circles represent the observation data from  \citet{Ilbert2013}. The black and red components in the top panel denote the total and quiescent galaxies, respectively. The error bars of number fraction and quiescent fraction indicate the $1\sigma$ range.
  }
	\label{fig:H15}
\end{figure}
 
 We define the massive galaxies(MGs) as the galaxies with the stellar masses $M_* \geq 10^{11}$M$_{\sun}$, and identify a galaxy as quiescent when its specific SFR (sSFR) satisfied $\textrm{sSFR} < \frac{1}{3t_{\textrm{H}(z)}}$, where ${t_{\textrm{H}(z)}}$ is the Hubble time at the redshift $z$ of the galaxy. Otherwise, the galaxies are regarded as star-forming ones. Fig.~\ref{fig:H15} shows the number density (top panel) and the quiescent fraction (bottom panel) of MGs as a function of redshift. The observational data from \citet{Muzzin2013} and \citet{Ilbert2013} are presented by circles and square symbols individually, while the simulation result from \citetalias{Henriques2015} (H15) are presented as solid curves. The black and red indicate the total and quiescent galaxies, respectively. To consider the magnitude bias effect in the simulated sample, the stellar mass of each simulated galaxy is convolved with a Gaussian in log$M_{*}$, with width 0.08 $\times (1 + z)$.  
 
 We find that the number density of all simulated MGs increases with decreasing redshifts, by about $2$ orders from $z=4$ to $0$. While the number density of the quiescent simulated MGs increases more drastically. The observed number density of all MGs and quiescent MGs are a slightly more than our simulated results but within 3 sigma confidence level. The difference becomes bigger as the redshift increases. This possibly indicates that the quenching mechanism should have a weaker dependence on the cosmic time in the SAM model. This should be investigated carefully in the development of the models next time. Although there is a divergence in the number density between the simulation and observational results, the predicted quiescent fraction coincide with the observations very well.
 
 The quiescent fraction increase from a few per cent at $z=4$ to about $80\%$ at $z=0$. This ensures we could study the quenching process of MGs in the following sections.

\section{Formation and Evolution of Massive Galaxies}

In this section, we will check the formation and evolution of MGs. Firstly we define a formation time of an MG, $t_{\rm form}$, as the cosmic time when its stellar mass grows to 10$^{11} $M$_{\sun}$ for the first time; the corresponding redshift as its formation redshift $z_{\rm form}$. We will study the different evolution phases of MGs: before and after their $t_{\rm form}$, respectively.

\subsection{The formation of MGs}

\begin{figure}
	\includegraphics[width=\columnwidth]{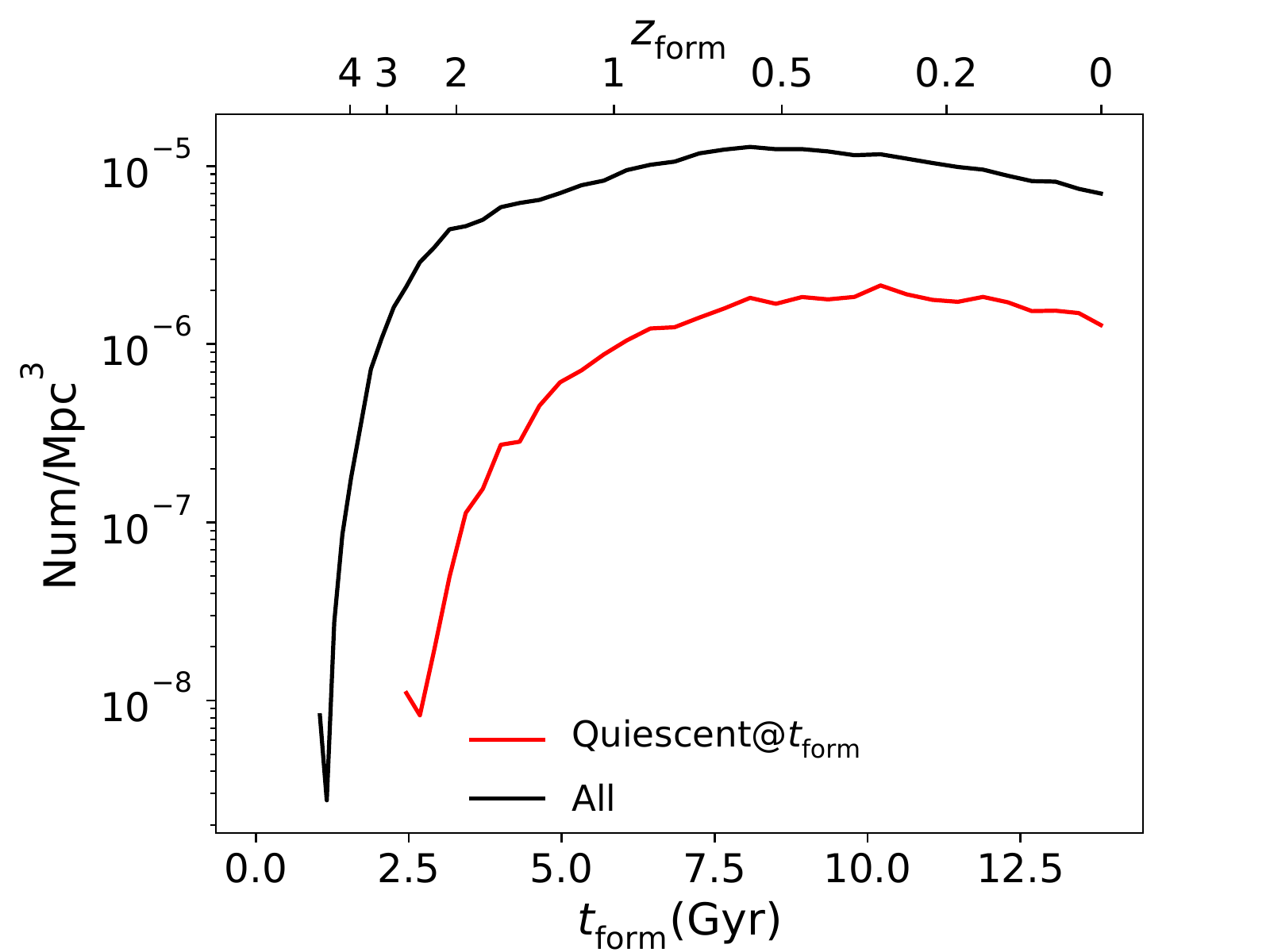}
	\caption{The distribution of formation time of new forming MGs at ($t_{\rm form}$ which is defined as the cosmic time when a galaxy's stellar mass firstly grows up to 10$^{11} $M$_{\sun}$). The black and red lines denote the number density of all MGs and the quiescent sub-sample at $t_{\rm form}$, respectively. The $z_{\rm form}$ is also labelled in the top x-axis.}
	\label{fig:form1}
\end{figure}

In Fig.~\ref{fig:form1}, we checked the formation time of the MGs. The number density of new forming MGs at each cosmic time is presented as the black solid curve.  While the number density of those MGs already being quiescent at $t_{\rm form}$ is presented as the red solid curve. A few  MGs form even earlier than $z=5$, though their number density is less than $\sim 10^{-8}$ $\textrm{Mpc}^{-3}$. The number density of MGs increases (from $\sim 10^{-8}$ to $\sim 10^{-5}$) rapidly from $z_{\rm form} \sim6$ to $z_{\rm form} \sim2$, then slowly climbs to the peak around $z_{\rm form}=0.6$. The MGs which have been quiescent at $t_{\rm form}$ emerge for the first time at $z\sim2.5$. Note that, in Fig.~\ref{fig:H15}, the ``quiescent MGs'' already exist at $z\sim 3.5$. Those galaxies are not quiescent at their forming time but became quiescent later around $z\sim 3.5$. Different from the high fraction of quiescent MGs at the low redshifts as shown in Fig.~\ref{fig:H15}, we find that the fraction of quiescent MG at the formation time is always low ($\sim 0.3$\% at $z\sim 2.5$, and $\sim 15$\% at $z<1$), and only $\sim13$\%, including MGs with all $z_{\rm form}$.

\begin{figure}
	\includegraphics[width=\columnwidth]{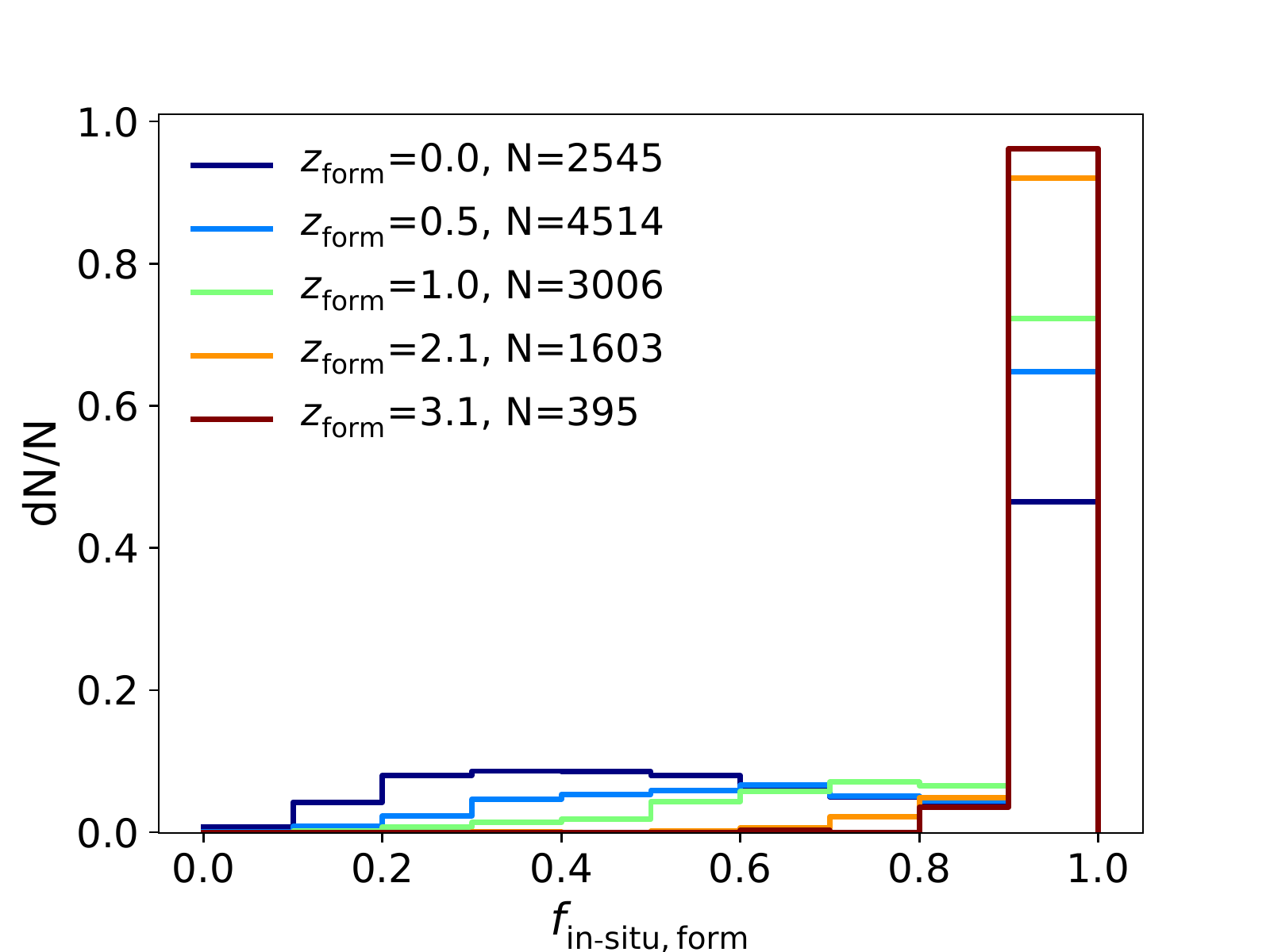}
	\caption{The distribution of $f_{\rm in\mhyphen situ,form}$ (in-situ mass fraction at $z_{\rm form}$) for MGs with different $z_{\rm form}$.}
	\label{fig:form3}
\end{figure}

\begin{figure}
	\includegraphics[width=\columnwidth]{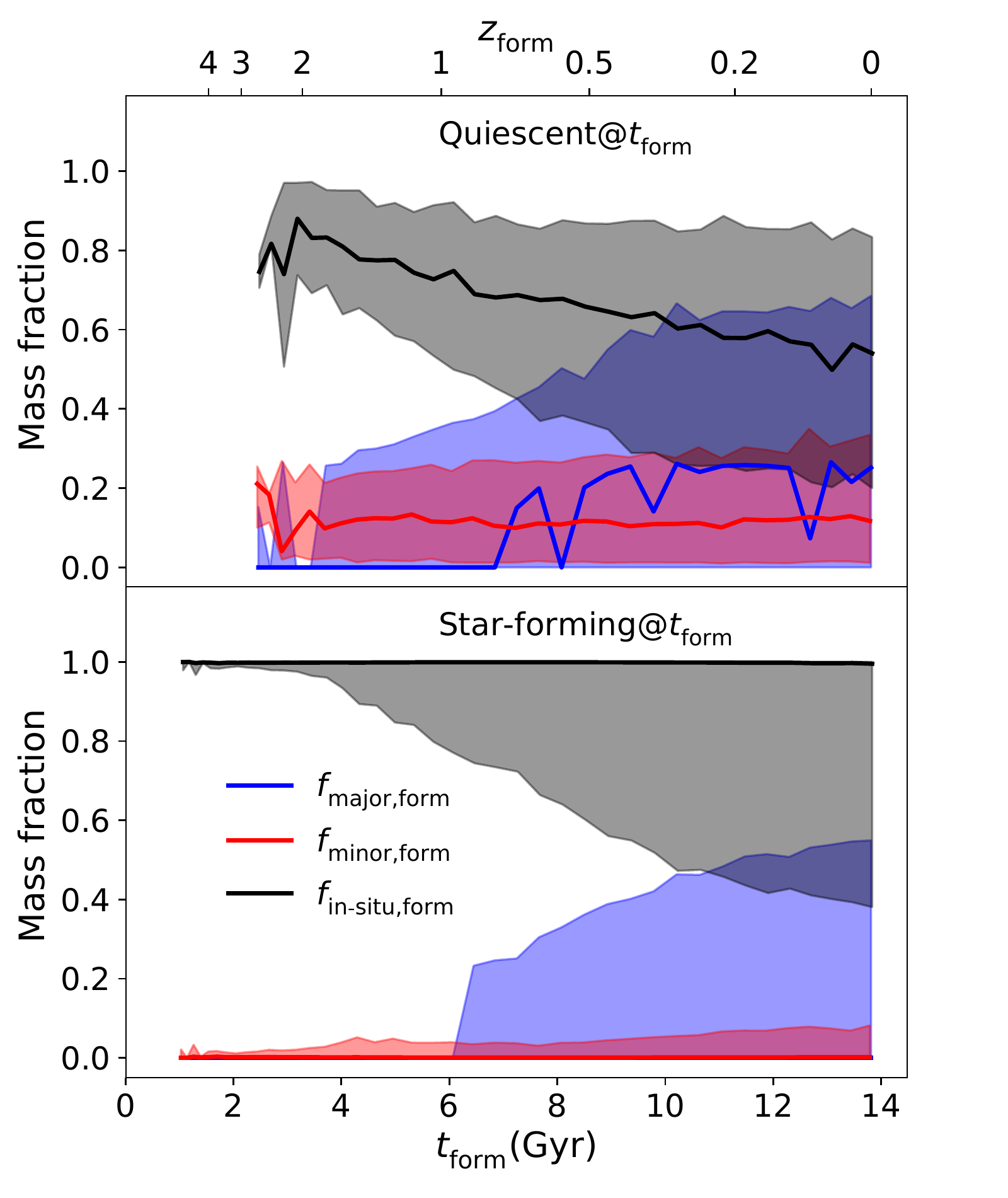}
	\caption{Mass fraction of stellar components from different processes as a function of $t_{\rm form}$ for MGs been quiescent (top panel) and star-forming (bottom panel) at $z_{\rm form}$. The black, blue and red solid lines represent the median mass fraction of in-situ ($f_{\rm in\mhyphen situ,form}$), major merger ($f_{\rm major,form}$) and minor merger ($f_{\rm minor,form}$), respectively. The shaded regions correspond to the range between the 16th and 84th percentiles.}
	\label{fig:form4}
\end{figure}

It will be interesting to understand how an MG assembles its stellar components in such a short time (noting the formation time of some MGs is kind of $\sim 1$Gyr) and why a fraction of them gets quiescent at their formation time. In the following, we study the stellar mass assembly of MGs before their $t_{\rm form}$. In the hierarchical formation scenario, the stellar mass of a galaxy can assemble through in-situ star formation or ex-situ \citep[i.e., mergers; e.g.][]{Guo2008,Oser2010,Lackner2012,Rodriguez-Gomez2016,Qu2016}.

The galaxy merger can also be classified as a major merger and minor merger according to the stellar mass ratio, $M_1/M_2$, of the main progenitor to the sum of its companions in a merging system. We classify $M_1/M_2<3$ as the major mergers, while $M_1/M_2>3$ as the minor mergers \citep[e.g.][]{Cole2000, DeLucia2007, Guo2008,Xie2015}.
For each MG, we calculate the fraction of stellar mass assembly from the three channels: major mergers ($f_{\rm major,form}=M_{\rm{major}}/M_{\rm{form}}$), minor mergers ($f_{\rm minor,form}=M_{\rm{minor}}/M_{\rm{form}}$) and in-situ star formation ($f_{\rm in\mhyphen situ,form}=M_{\rm{in\mhyphen situ}}/M_{\rm{form}}$), where $M_{\rm{major}}$, $M_{\rm{minor}}$, and $M_{\rm{in\mhyphen situ}}$ denotes the total stellar mass formed in major mergers, minor mergers, and in-situ star formation before $t_{\rm form}$, respectively, and $M_{\rm{form}}$ shows the stellar mass of an MG at $t_{\rm{form}}$ ($M_{\rm{form}}$ is slightly larger than  $10^{11}\ M_{\odot}$ because $t_{\rm{form}}$ is defined as the redshift of the first snapshot after the mass of MG is greater than $10^{11}\ M_{\odot}$.)

In Fig.~\ref{fig:form3}, we show the distributions of $f_{\rm in\mhyphen situ, form}$ of the MGs at the different redshifts with different colours. And the number of MGs in each bin is given in the legend. 
We find that, in statistics, the in-situ star formation dominates the stellar mass assembly of the MGs at all redshifts. At higher redshift $z_{\textrm{form}}=3.1$, about $95\%$  MGs assemble their stellar mass by in-situ way  ($f_{\rm in\mhyphen situ,form} \sim 1$ ). This fraction is still as high as $45\%$ at redshift zero. With the decreasing redshift, the mass contribution from mergers is becoming more important for the formation of the MGs. At $z=0$, the contribution from in-situ way decreases down to less than 50 per cent and the star formation is dominated by galaxy mergers. 

The existence of quiescent MGs at $t_{\rm form}$  poses an interesting question: what processes quench them after they grow quickly to be MGs at high redshift?  So we subsequently study the stellar mass assembly of MGs being quiescent (MG-FQs) and star-forming (MG-FSFs) at $t_{\rm form}$, respectively. Fig.~\ref{fig:form4} shows the fractions of the stellar masses assembled from the major, minor mergers, and in-situ star formation as a function of $t_{\rm form}$, for the MG-FQs (top panel) and MG-FSFs (bottom panel), respectively. The black, blue and red curves represent the median $f_{\rm in\mhyphen situ,form}$, $f_{\rm major,form}$, and $f_{\rm minor,form}$, respectively. The coloured shaded regions bracket the $1\sigma$ fraction ranges (from 16\% to 84\%).  Most of the MG-FSFs assemble their stars mostly through the in-situ way and almost have undergone neither major nor minor mergers.  While for MG-FQs, the in-situ way is still dominating for the stellar assembly, but the minor mergers constantly contribute about more than 10\% at any $t_{\rm form}$. At $z_{\rm form}<1$, more than 20\% have major mergers before they form, and the younger MG-FSFs are more likely to have undergone major mergers. This indicates, although the in-situ dominate the growth of the MGs, the mergers, especially the major mergers, are possibly responsible for the quenching of MGs. These results are in qualitative agreement with results from the Illustris simulation \citep[][fig. 6]{Rodriguez-Gomez2016}, as well as EAGLE simulation \citep[][fig. 6]{Qu2016}.

\begin{figure}
    \centering
	\includegraphics[width=1\columnwidth]{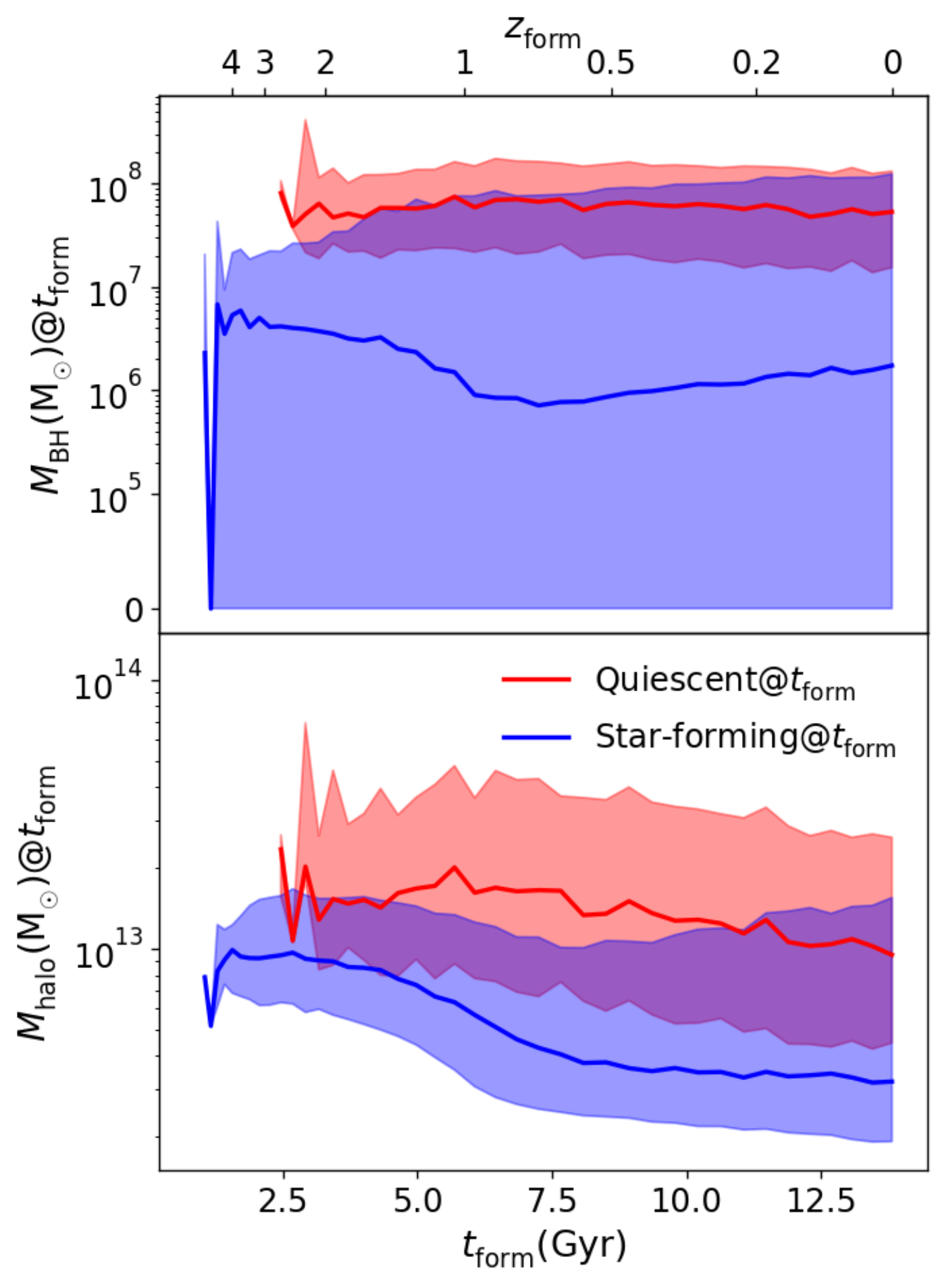}
	\caption{The central black hole mass (top panel) and halo mass (bottom panel) of MGs at $t_{\rm form}$ as a function of their $t_{\rm form}$. The blue and red lines denote the median values of the MGs which are star-forming and quiescent at $t_{\rm form}$, respectively. The shaded regions correspond to the range between the 16th and 84th percentiles}
	\label{fig:form2}
\end{figure}

In the \citetalias{Henriques2015} model, galaxy mergers can significantly enhance the growth of central black holes (BHs) in the galaxies, then trigger the AGN activity and feedback \citep{Kauffmann2000,Benson2003,Granato2004,Springel2005,Bower2006,Croton2006}. AGN feedback can suppress gas cooling and prevent star formation.  In \citetalias{Henriques2015}, the power of AGN feedback is positively related to the BH mass and halo mass.

We therefore present the BH mass ($M_{\rm BH}(t_{\rm{form}})$, top panel) and halo mass ($M_{\rm halo}(t_{\rm{form}})$, bottom panel) of the MGs as a function of $t_{\rm form}$, as shown in Fig.~\ref{fig:form2}.
The blue and red curves highlight the median values of MG-FSFs and MG-FQs at each $t_{\rm form}$, respectively. The coloured shaded components bracket the $1\sigma$ mass ranges.

For the MG-FSFs and MG-FQs with the same $t_{\rm form}$, the BH and halo masses of the MG-BQs are more massive than those of the MG-FSFs. It indicates the fact that the MG-FQs have undergone more galaxy mergers, compared with the star-forming counterparts, and more mergers trigger the faster growth of the central BHs of galaxies and reproduce stronger AGN feedback which leads to a quicker suppression of star formation activity. This result agrees with that of \citet{Terrazas2016}, in which the authors compared the central BH masses of the quiescent and star-forming galaxies with the similar stellar masses in the nearby universe ($z < 0.034$), and found that the quiescent galaxies host more massive BHs.

We also find that the $M_{\rm BH}(t_{\rm form})$ of the MG-FQs are always about $10^{7.5}$ M$_{\sun}$, independent of the different $t_{\rm form}$. The value is higher than the $M_{\rm BH}$ ($\sim$ $10^6\--10^7$ M$_{\sun}$) of the star-forming counterparts. It implies that a BH with $M_{\rm BH}$ $\sim$ $10^7$ M$_{\sun}$ is not massive enough to quench MGs. However, this characteristic value $M_{c}$ depends on the chosen stellar-mass threshold $M_{thre}$ in defining the MGs. The dependence approximates $Log M_{c} \propto Log M_{thre}$ and agrees with the earlier found black hole mass -- stellar mass relation \citep[e.g.][]{Malbon2007,Somerville2008,Terrazas2020}. This indicates the strong co-evolution of the black holes and its host galaxies: the strength to quench the galaxies depends on the strength of the AGN feedback, and as well on the mass of the black hole. 

From the bottom panel of Fig.~\ref{fig:form2}, we find that the MGs with earlier $t_{\rm form}$ tend to be hosted in the more massive halos, which is in agreement with the evolution of the stellar mass \-- halo mass (SMHM) relation in \citet{Matthee2017}. They find that the ratio between stellar mass and halo mass decrease with increasing redshift at $z>0.3$.

\subsection{The evolution of MGs} \label{ssec:afterform}

\begin{figure}
	\includegraphics[width=\columnwidth]{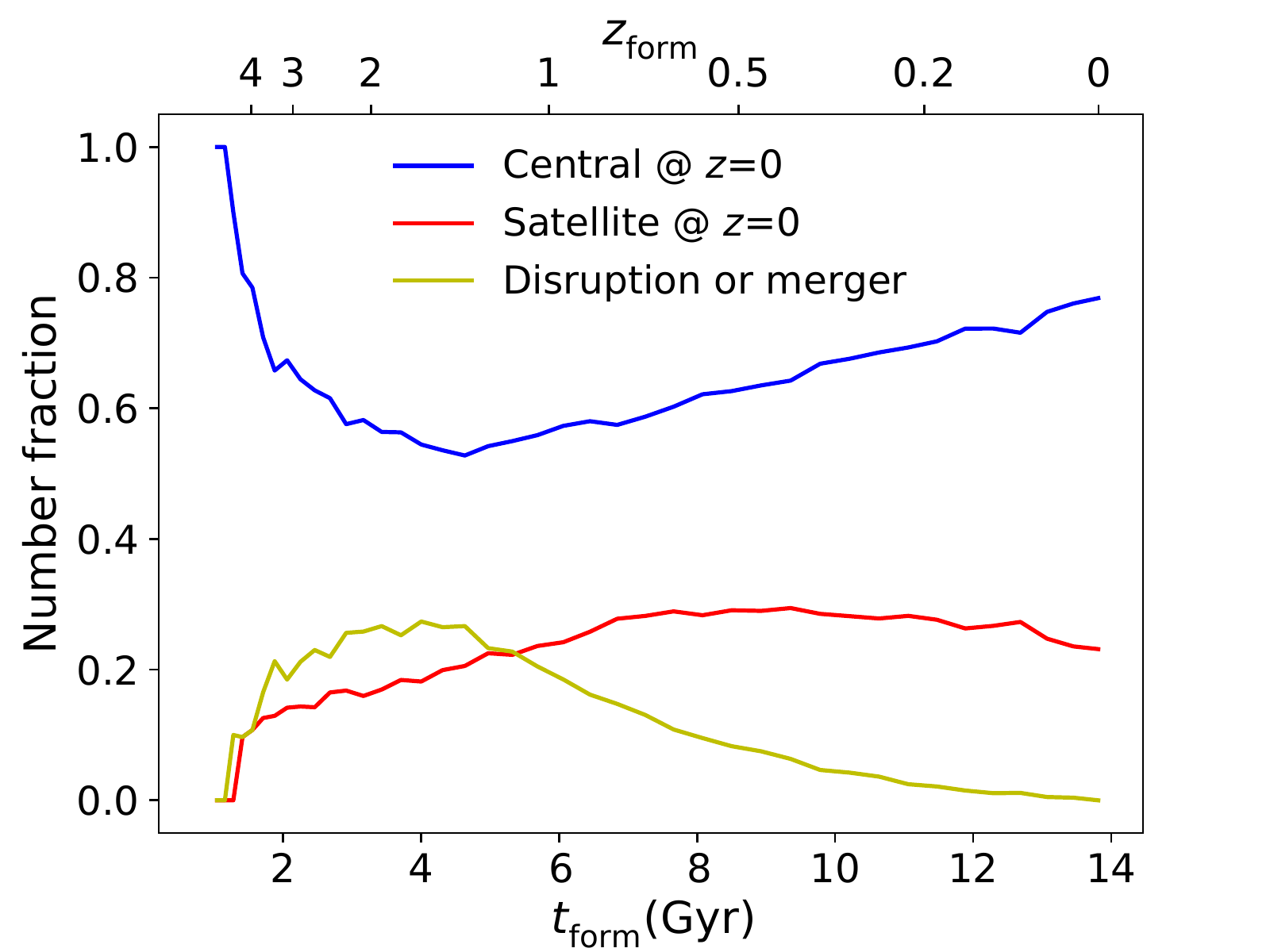}
	\caption{The number fraction of MGs with different fate at $z=0$ as a function of $t_{\rm form}$. The blue and red lines represent the number fraction of MGs evolved to $z=0$ as central or satellite, respectively. The fraction of galaxies disrupted or merged with other more massive galaxies before $z=0$ is denoted by the yellow line.}
	\label{fig:aft1}
\end{figure}

In this subsection, we investigate the evolution of MGs after $t_{\rm form}$. 
In \citetalias{Henriques2015}, a galaxy initially emerges at the centre of a halo, named the ``central galaxy'', and the halo (together with the galaxy) may fall into another halo with a deeper gravitational potential, becoming a subhalo orbiting within the FOF group, then the galaxy is named as the ``satellite galaxy''. A satellite may merge into the central galaxy after an episode of dynamical time from infall. During this time, the satellite may be disrupted by tidal force and contribute to the intra-cluster light \citep{Guo2011}.

Therefore, an MG may have three different fates at $z=0$, i.e. being a central, or a satellite, or vanished (i.e., have been disrupted or have merged into other more massive galaxies before $z=0$). For the MGs with the different $t_{\rm form}$, Fig.~\ref{fig:aft1} shows the number fraction of the MGs ends with three different fates at $z=0$. The blue and red colours highlight the number fraction of the MGs which are the central and satellite galaxies at $z=0$, respectively. The fraction of galaxies that have been vanished before $z=0$ is shown by the yellow colour. 
 
For the MGs with $z_{\rm form}$ $>1.5$, the fraction of MGs still being central at $z=0$ decreases with the decreasing $z_{\rm form}$, and a fraction of MGs vanished before $z=0$ increases with decreasing $z_{\rm form}$, which is due to the fact that MGs with earlier $t_{\rm form}$ tend to be hosted in more massive halos (bottom panel of Fig.~\ref{fig:form2}) and have deeper gravitational potential. Therefore, they are more likely to cannibalize the other satellite galaxies, rather than fall into another halo potential and become satellites.
However, for the MGs with $z_{\rm form}<1.5$, the fraction of MGs still being central at $z=0$ increases with the decreasing $z_{\rm form}$, while the fraction of the vanished MGs before $z=0$ decreases with the decreasing $z_{\rm form}$. It is primarily because that the number of new-formation MGs is approximately uniform when $z_{\rm form}<1.5$, as shown in Fig.~\ref{fig:form1}, and therefore, the MGs with later $t_{\rm form}$ do not have enough time to merge into other systems or be disrupted.

Note, the total number fraction of MGs which are vanished before $z=0$ accounts for $\lesssim 11\%$ of the MGs sample. Therefore, we will only study the evolution of the survived MGs at $z=0$ hereafter.

\begin{figure}
	\includegraphics[width=\columnwidth]{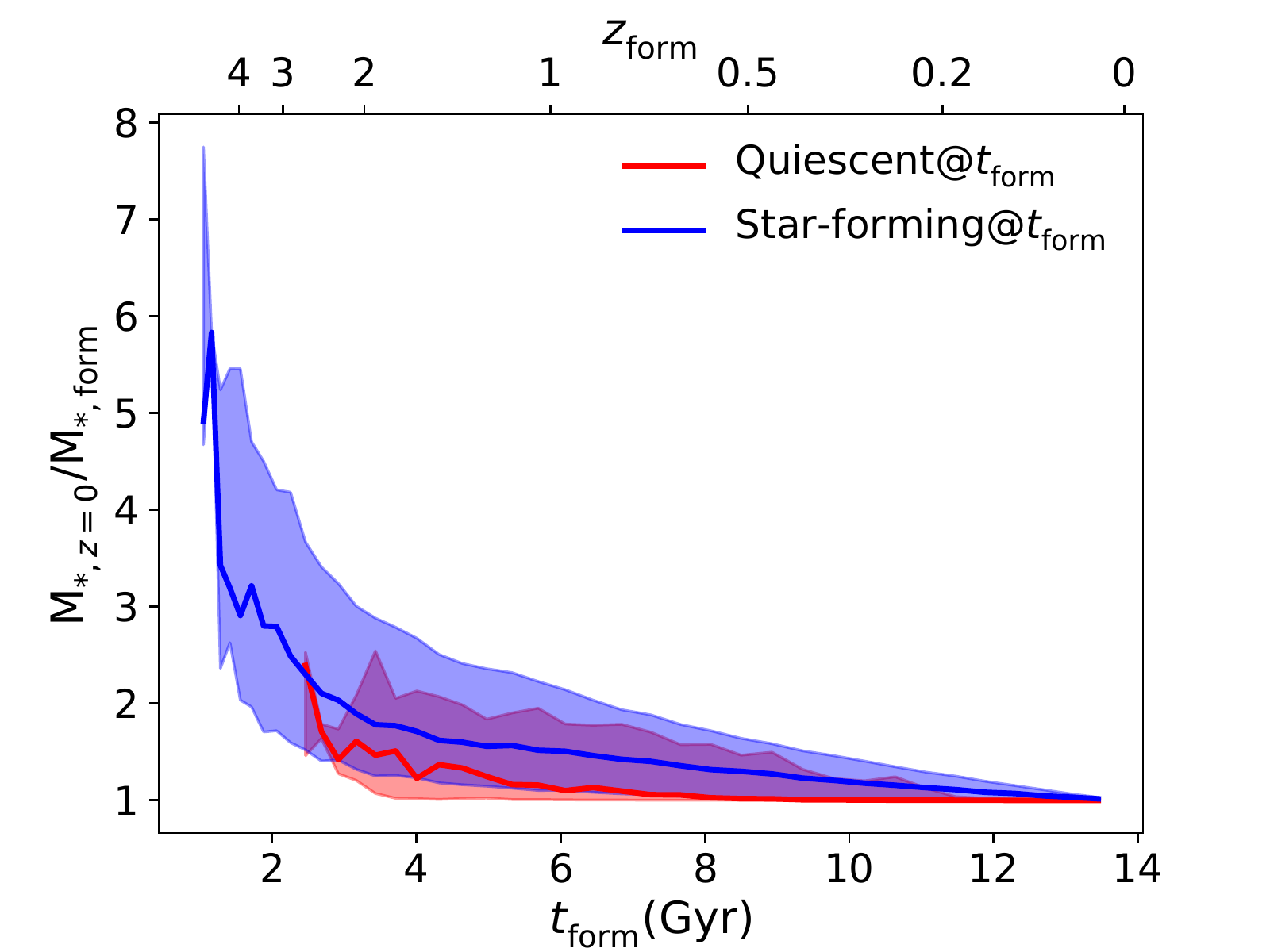}
	\caption{The stellar mass mass ratios of MGs at  $z=0$ to $t_{\rm form}$ as a function of $t_{\rm form}$ for MGs been quiescent (red) and star-forming (blue) at $t_{\rm form}$ . The solid lines denote the median value and the shaded regions represent the range between the 16th and 84th percentiles. }
	\label{fig:aft3}
\end{figure}

\begin{figure}
	\includegraphics[width=\columnwidth]{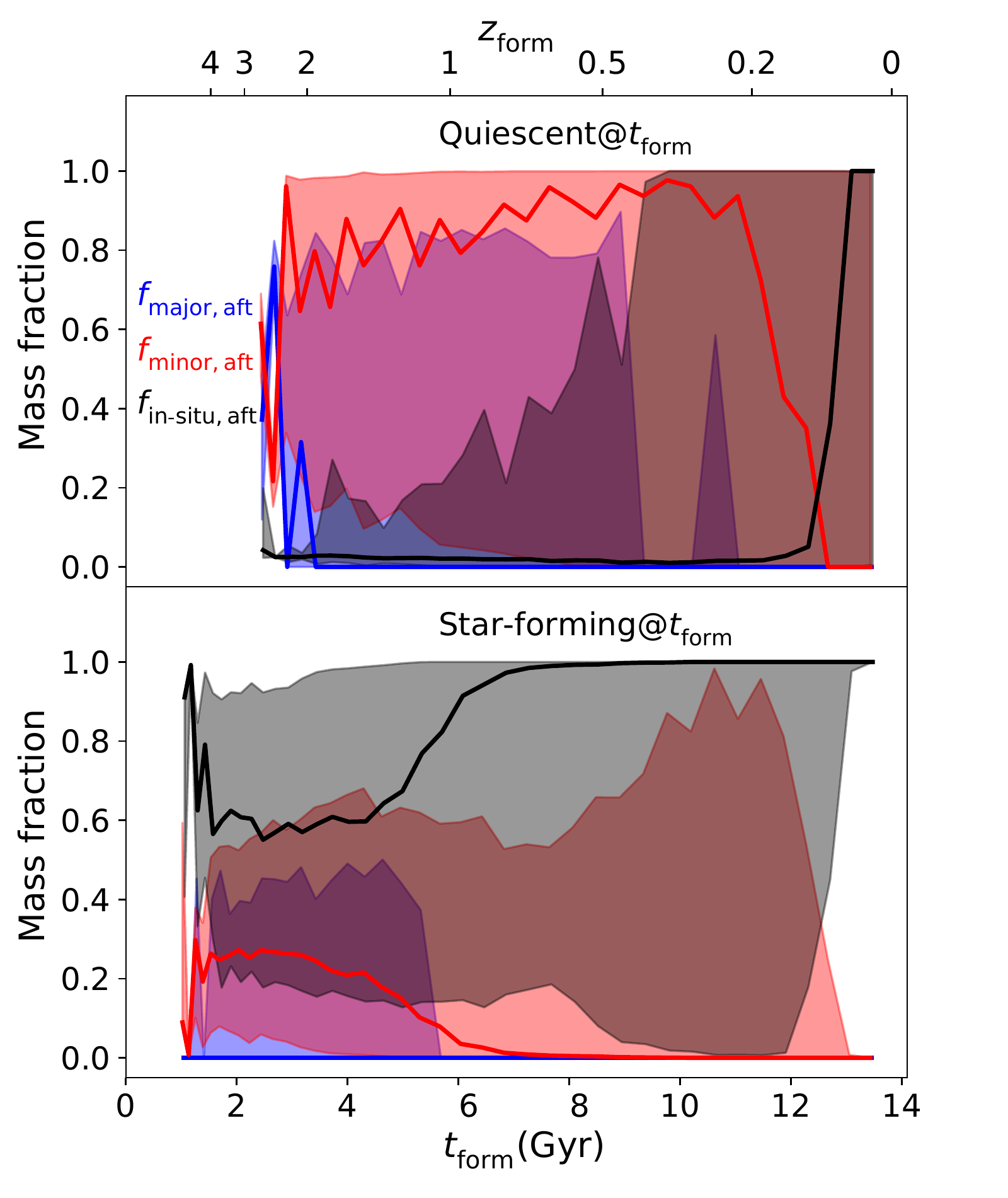}
	\caption{Stellar mass fraction from different growth processes after $t_{\rm form}$ until $z=0$ for MGs been quiescent (top panel) and star-forming (bottom panel) at $t_{\rm form}$. The black, blue and red solid lines represent the median mass fraction of in-situ ($f_{\rm in\mhyphen situ,aft}$), major merger ($f_{\rm major,aft}$) and minor merger ($f_{\rm minor,aft}$) growth process, respectively. The shaded regions correspond to the range between the 16th and 84th percentiles.}
	\label{fig:aft4}
\end{figure}

First, we investigate how many stellar masses the MGs can assemble from $z_{\rm{form}}$ to $z=0$. In Fig.~\ref{fig:aft3}, we plot their ratios of the stellar masses at $z=0$ to $M_{\rm{form}}$, as a function of $t_{\rm form}$. The blue and red components show the median value and $1\sigma$ scatter of the ratios of MG-FSFs and MG-FQs, respectively. We find that the MGs only grows about a factor of a few even some of them formed at $z=4$. 
We also find that the stellar mass growth of MG-FSFs is more significant, compared with the MG-FQs with the same $t_{\rm form}$. 

Subsequently, for each MG, we calculate the fraction of the stellar mass assembly through major mergers ($f_{\rm major,aft}=M_{\rm{major,aft}}/(M_{z=0}-M_{\rm{form}})$), minor mergers ($f_{\rm minor,aft}=M_{\rm{minor,aft}}/(M_{z=0}-M_{\rm{form}})$), and in-situ star formation ($f_{\rm in\mhyphen situ,aft}=M_{\rm{in\mhyphen situ,aft}}/(M_{z=0}-M_{\rm{form}})$), respectively; the $M_{\rm{major,aft}}$, $M_{\rm{minor,aft}}$, and $M_{\rm{in\mhyphen situ,aft}}$ denotes the total stellar mass formed in major mergers, minor mergers, and in-situ star formation after $t_{\rm form}$, respectively, and the $M_{z=0}$ denotes the stellar mass of the MG at $z=0$.

In Fig.~\ref{fig:aft4}, we present the stellar mass assembly fractions as a function of $t_{\rm form}$ for the MG-FQs (upper panel) and MG-FSFs (lower panel), respectively. The black, blue and red components respectively represent the median $f_{\rm in\mhyphen situ,aft}$, $f_{\rm major,aft}$ and $f_{\rm minor,aft}$, and their $1\sigma$ scatters.
For the MG-FQs, unlike the stellar mass growth before $t_{\rm{form}}$, the stellar mass growth after $t_{\rm{form}}$ is dominated by the minor mergers. For some MG-FQs form close to $z=0$, the in-situ way plays the key role although the mass growth is tiny. 

For the MG-FSFs, in statistics, the in-situ star formation is responsible for their stellar mass growth after $t_{\rm form}$, particularly for the MG-FSFs with $z_{\rm{form}}<1$. This result is consistent with the conclusion of Fig.~7 which we mentioned before. While for MG-FSFs with  $z_{\rm form} \gtrsim1.5$, the minor merge is also important in their stellar mass assembly after $z>z_{\rm{form}}$. 

\begin{figure}
	\includegraphics[width=\columnwidth]{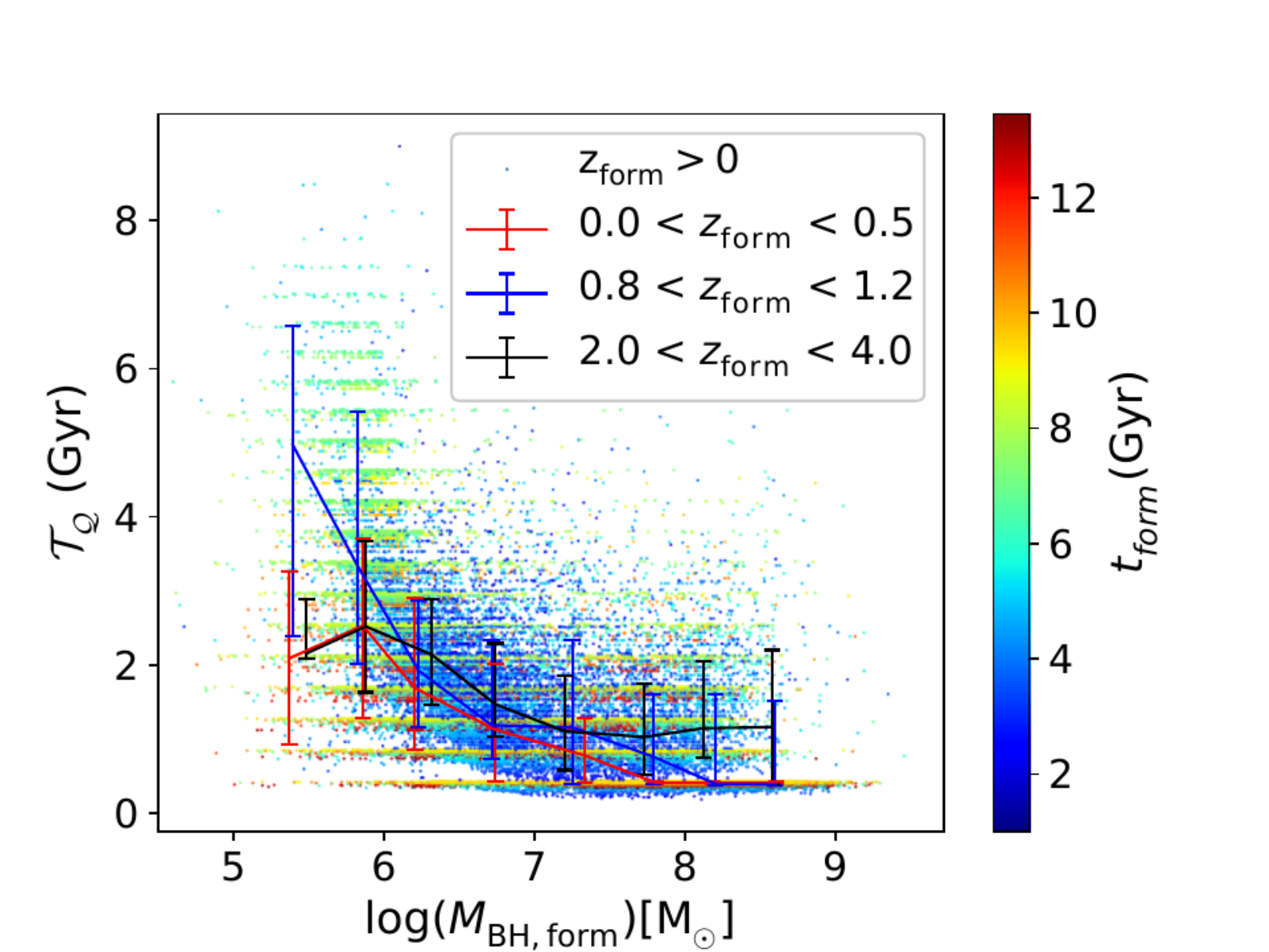}
	\caption{Scatter plot of quenching timescales, $\mathcal{T}_{\mathcal{Q}}$ vs. $M_{\rm BH,form}$, the black hole mass of MGs when they form. The colours of points code the $t_{\rm form}$ of MGs. The red, blue and black lines represent the median value of $\mathcal{T}_{\mathcal{Q}}$ in each log($M_{\rm BH,form}$) bin for MGs with $0<z_{\rm form}<0.5$, $0.8<z_{\rm form}<1.2$ and $2<z_{\rm form}<4$, respectively. The error bars represent the $16\--84$th percentiles ranges.}
	\label{fig:aft2}
\end{figure}

As shown in Fig.~\ref{fig:form1}, only a small fraction of MGs has been quenched when they formed. However, most of the MGs at $z=0$ lack star formation activity. In order to study the quenching mechanism of MG-FSFs after $t_{\rm form}$, for each MG-FSF, we trace its evolution history after $t_{\rm form}$ and define the quenching timescale, $\mathcal{T}_{\mathcal{Q}}$, as the interval between $t_{\rm form}$ and the critical time $t_{\rm c}$ of sSFR$\lesssim 1/3t_{\textrm{H}(z)}$. Note, a quiescent galaxy may resume a large sSFR$>1/3t_{\textrm{H}(z)}$ during a later gas-rich merger; in this case, $t_{\rm c}$ denotes the first time its sSFR decreases to the critical value $\sim 1/3t_{\textrm{H}(z)}$. 

As Fig.~\ref{fig:form2} indicates, the supermassive black hole in the MGs may be responsible for their quenching.
In Fig.~\ref{fig:aft2}, we checked the dependence of $\mathcal{T}_{\mathcal{Q}}$ on their black hole mass $M_{\rm BH,form}$ at $t_{\rm form}$. Every data point presents a galaxy, and its colour labels its $t_{\rm form}$. The red, blue and black lines represent the median value of $\mathcal{T}_{\mathcal{Q}}$ in each log($M_{\rm BH,form}$) bin for MGs with $0<z_{\rm form}<0.5$, $0.8<z_{\rm form}<1.2$ and $2<z_{\rm form}<4$, respectively. The error bars are for  the $1\sigma$ ranges. The MGs whose sSFR never decreases to less than $1/3t_{\textrm{H}(z)}$ and MGs with no black hole at $t_{\rm form}$ are not shown in this figure. We find that the $\mathcal{T}_{\mathcal{Q}}$ has noticeable dependence on $M_{\rm BH,form}$ for MGs forms at different $t_{\rm form}$. The $\mathcal{T}_{\mathcal{Q}}$ decrease sharply with increasing $M_{\rm BH,form}$ until $M_{\rm BH,form}\sim10^{7.5}$ M$_{\sun}$, and then almost remains as a very small time scale. The median value of the $\mathcal{T}_{\mathcal{Q}}$ of MGs with $M_{\rm BH,form}>10^{7.5}$ M$_{\sun}$ is only $\sim 0.4$ Gyr. About 45\% of MGs with no black hole at $t_{\rm form}$ (not shown) are quenched before $z=0$ and the median value of their $\mathcal{T}_{\mathcal{Q}}$ is $\sim3.3$ Gyr, roughly equal to $\mathcal{T}_{\mathcal{Q}}$ of MGs with lowest $M_{\rm BH,form}$ ($\sim 10^5$ M$_{\sun}$). These results indicates that the balck halo with mass  $M_{\rm BH,form}\sim10^{7.5}$ M$_{\sun}$ drives the quenching process of the MGs by their strong AGN feedback which are more effectively to suppress cooling and star formation. However, the characteristic value of the $M_{\rm BH,form}$ depend on our chosen stellar mass threshold in MGs definition, which reflects the strong relation between the black hole mass and the mass of their host galaxies.

\begin{figure*}
	\includegraphics[width=2\columnwidth]{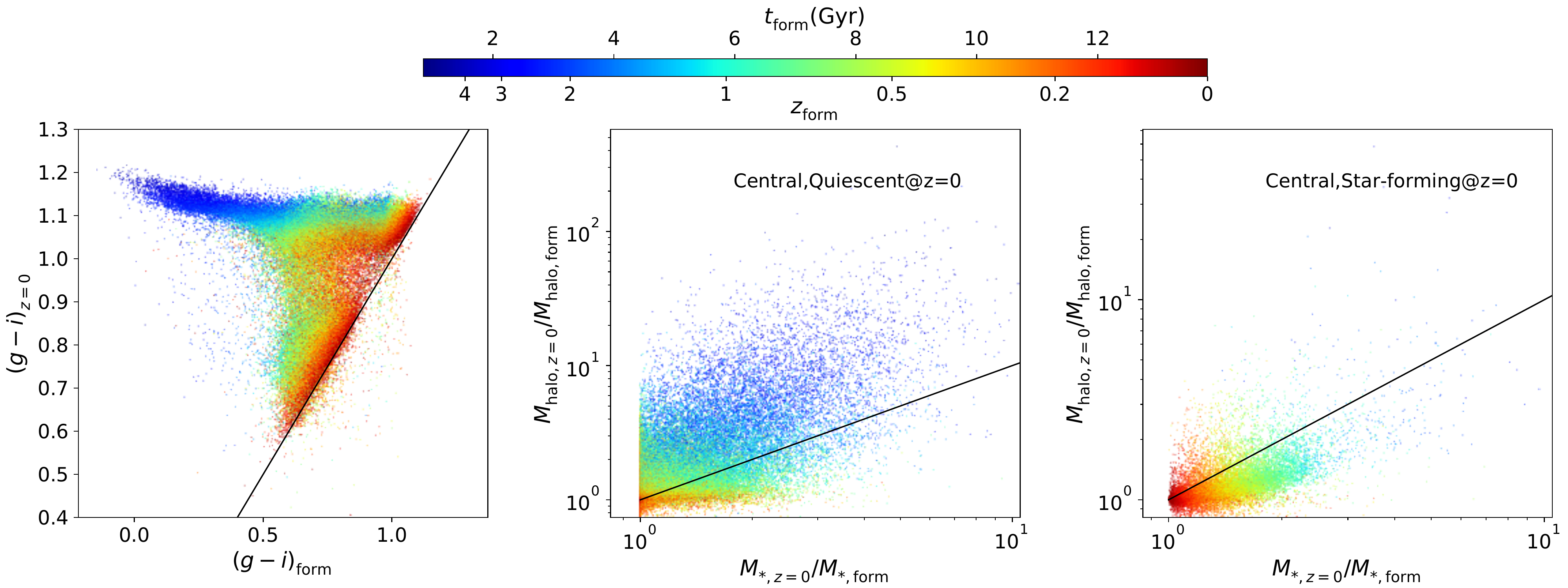}
	\caption{Scatter plot of $g-i$ at $z=0$ vs. $g-i$ at $t_{\rm form}$ (left panel) and stellar mass ratio of MGs at $z=0$ to $z_{\rm form}$ vs. halo mass ratio of MGs at $z=0$ to $z_{\rm form}$ for the central MGs being quiescent (middle panel) and star-forming (right panel) at $z=0$. The colours of points represent the $t_{\rm form}$ of MGs. The black lines in the three panels represent where the values of the x-axis equal to the y-axis.}
	\label{fig:aft5}
\end{figure*}

It is also interesting to check the evolution of the MGs's colours from their formation time until redshift $z=0$. The left panel of Fig.~\ref{fig:aft5} shows the scatter plot of $g-i$ at $z=0$ vs. $g-i$ at $t_{\rm form}$ for MGs with different $t_{\rm form}$. Almost all of MGs with $t_{\rm form} \gtrsim 1$ have changed into red at $z=0$. For those MGs formed after $z=0.2$, most of them don't change too much, but some of them turn red soon, indicating they just suffered a very quick quenching. More than half of these quickly quenched galaxies host a BH with a mass larger than  $10^{7.5}$M$_{\odot}$, and the later could be responsible for the quenching. 

In the middle and right panels of Fig~\ref{fig:aft5}, we check the co-evolution of these MGs and their host dark matter haloes by showing the increase of the mass of host dark matter halo vs. the increase of their stellar mass. In the right panel, for those MGs which are still in the star-forming phase at redshift zero, their stellar mass growth is even a little faster than the mass growth of their host dark matter halo; while for those quiescent MGs at redshift zero in the middle panel, the mass growth of the host dark matter halo is quicker than stellar mass, especially for the MGs formed before $z=1$, dark halo mass of some is almost 10 times faster than the growth of stellar component. 

\section{The properties of Massive Galaxies with redshift}
\begin{figure*}
	\includegraphics[width=2\columnwidth]{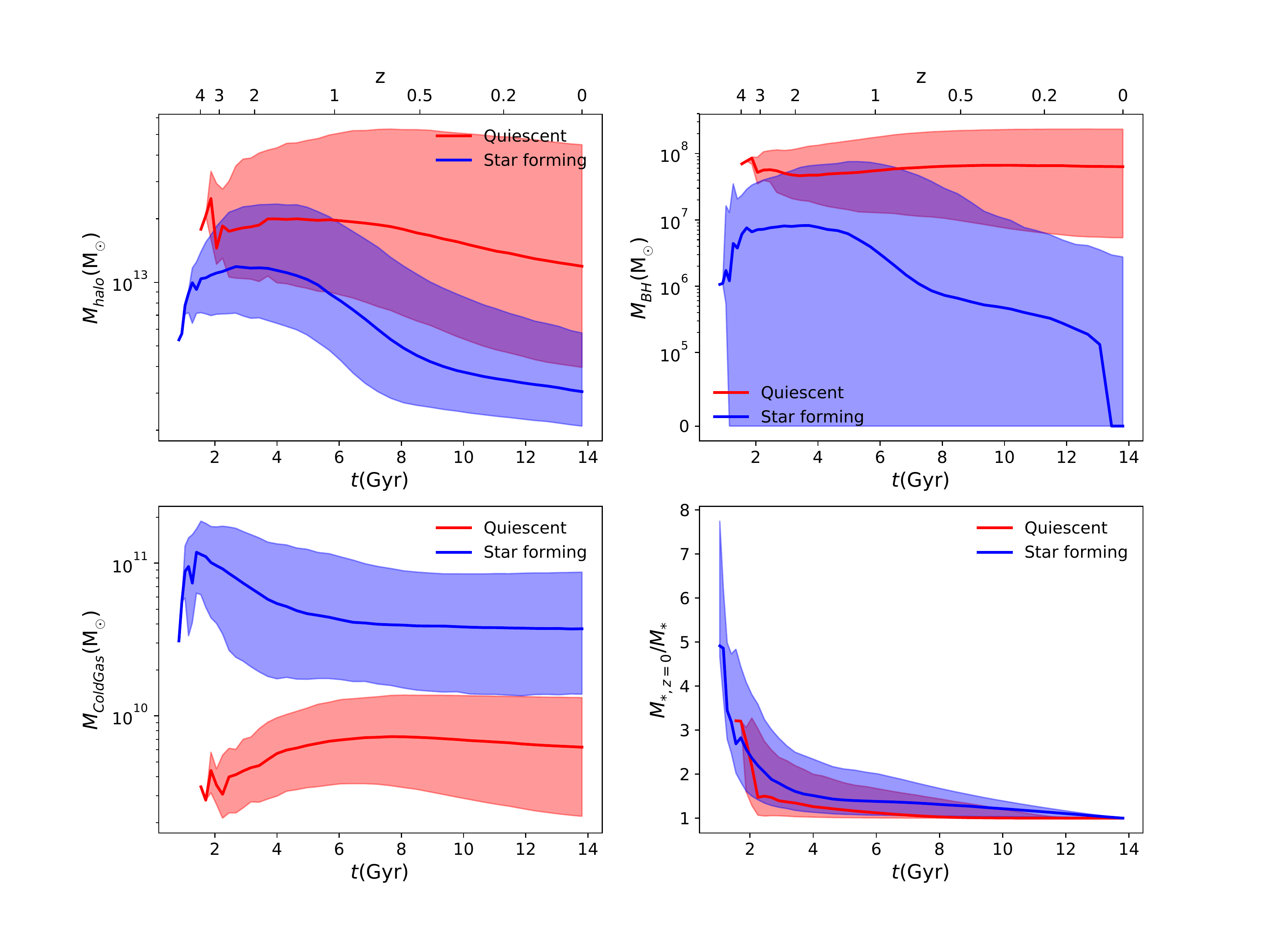}
	\caption{The halo mass (upper left panel), black hole mass (upper right panel), cold gas mass (bottom left panel) and stellar mass ratios of their descendant at $z=0$ and their own (bottom right panel) as a function of cosmic time for quiescent (red) and star-forming (blue) MGs. The solid lines represent the median values and the 16th and the 84th percentiles are represented by the shaded regions.}
	\label{fig:prop1}
\end{figure*}
In this section, we will focus our attention on the statistical properties of MGs at different redshifts. The upper left, upper right and bottom left panels of Fig.~\ref{fig:prop1} present the halo mass, black hole mass and cold gas mass as a function of the cosmic time for quiescent (red) and star-forming (blue) MGs, respectively. The solid lines denote the median value and the shaded regions for the range from the 16th to 84th percentiles. As shown in the figure, the properties of these two populations are obviously different. The quiescent MGs tend to stay in a more massive halo, with a larger black hole mass and less cold gas than star-forming MGs at all redshifts. The median value of the halo mass significantly decreases with the increase of cosmic time. This indicates that MGs prefer to stay in more massive haloes at high redshifts, and at low redshift, the MGs formed in smaller haloes becomes possible, especially for the star-forming MGs. 

The black hole mass of quiescent MGs, about thousandths of their stellar mass, have little change with redshifts, while the mass of BH in the star-forming MGs is almost always less than $10^7 M_{\sun}$ reflecting the very low AGN feedback inside. 
The bottom right panel of Fig.~\ref{fig:prop1} shows the stellar mass ratio of their descendant at $z=0$ and their own. Unlike the other three panels, the MGs not evolved to $z=0$ are not included in this panel.  We can find that the stellar mass assembly in the star formation MGs is a slightly quicker than the quiescent counterparts, but the difference is not so big given the former population is keeping to form star continuously.

\section{Discussion and Conclusions}
The formation and evolution of massive galaxies play an important role in understanding galaxies formation. In this paper, we have utilised the semi-analytic model developed by \citetalias{Henriques2015} to explore the formation and evolution of massive galaxies. When the stellar mass of a galaxy exceeds $10^{11}$M$_{\sun}$, we treat the galaxy as a massive galaxy (MGs) and its formation time defined by the comic time when its stellar mass grow to larger than $10^{11}$M$_{\sun}$ for the first time. We list here our principal findings:
\begin{enumerate}
	\item A few MGs have a very high $z_{\rm form}$ ($\sim6$). The number density of MGs increases from $z_{\rm form} \sim 6$ to $z_{\rm form} \sim 2$, and then almost remains unchanged until $z_{\rm form} = 0$. 
	\item There are only a small fraction (13\%) of MGs been quiescent at $t_{\rm form}$(MG-FQs). They have a more massive black hole and halo mass than the star-forming counterparts (MG-FSFs). We also find their black hole masses are always about $10^{7.5}$M$_{\sun}$, independent with their $t_{\rm form}$. However, this characteristic value depends on the chosen stellar-mass threshold in the definition of MGs, which hints at the co-evolution of the black hole and their host galaxies.
	\item We find that the in-situ star formation always dominates the stellar mass assembly at $t_{\rm form}$ although the mass factions of in-situ star formation are decreasing with increasing formation time. For the MG-FQs, although the in-situ star-formation still dominates the formation of their stellar masses, they are more likely to have undergone more galaxy mergers, compared with the formation scenario of MG-FSFs. These results are in qualitative agreement with results from the Illustris simulation \citep[][Fig. 6]{Rodriguez-Gomez2016}, as well as the EAGLE simulation \citep[][Fig. 6]{Qu2016}. In these simulations, the stellar mass fraction from the merger (``ex-situ") in MGs is $\sim 10\%$ to $\sim 30\%$, lower than ``in-situ". Our results also suggest the MGs has undergone very few major mergers before $t_{\rm form}$, but some other models argued more major merger events in the formation of MGs \citep[e.g.][]{DeLucia2007, Guo2008}. 
	This agrees with the fact found in \citep[][Fig. 2]{Yang2019}: the merger rates in \citetalias{Henriques2015} model is  relatively lower than other models.
	
	\item We find that about $25$\% of MGs sample have become satellite galaxies and $11$\% of MGs sample have been disrupted or merged into other more massive galaxies before $z=0$.
	\item For the MGs with at $1.5\lesssim z_{\rm form} \lesssim5$, their stellar mass assembly through mergers after $t_{\rm form}$ is comparable to that through in-situ star formation. But for the MGs with $z_{\rm form}\gtrsim5$, the in-situ star formation dominates their stellar mass assembly after $t_{\rm form}$.
	\item We find that the quenching timescales of MGs have a noticeable correlation with their $M_{\rm BH}$ and the correlation is not dependent on $t_{\rm form}$. The MGs with more massive $M_{\rm BH}$ have a shorter quenching timescale, suggesting the important effect of AGN feedback in the quenching of MGs.

\end{enumerate}
	
We noticed that a small fraction ($~6.4\%$)  of quenched MGs rejuvenated to becoming star-forming galaxies. Some of them will pass away again just less than 1Gyr, while a small fraction of them will keep star formation more than a few Gyr, and even keep alive at present. This rejuvenation phenomenon is also discovered in the observation \citep[e.g.][]{Thomas2010,Chauke2019,Cleland2021}. It is possible that the merger or some other gas supply mechanism is responsible for that. Next, we plan to carry on more further detailed studies on that in the model used in this study and also in the new state-of-the-art high-resolution hydrodynamic simulation.


\begin{acknowledgements}
	
We thank the referee for the valuable comments. We acknowledge supports from National Key R\&D Program of China (grant number 2018YFA0404503, 2018YFE0202902), the National Key Program for Science and Technology Research and Development of China (2017YFB0203300, 2015CB857005), the National Natural Science Foundation of China (Nos. 11988101, 11425312, 11503032, 11773032, 11390372, 11873051, 118513, 11573033, 11622325, 12033008, and 11622325). Y.R. acknowledges funding supports from FONDECYT Postdoctoral Fellowship Project No.~3190354 and NSFC grant No.\,11703037.
\end{acknowledgements}
  
\bibliographystyle{raa}
\bibliography{ms2021-0114}

\end{document}